\documentclass[english,3p]{elsarticle}
\usepackage[T1]{fontenc}
\usepackage[utf8]{inputenc}
\usepackage{float}
\usepackage{amsmath}
\usepackage{amsthm}
\usepackage{amssymb}
\usepackage{stmaryrd}
\usepackage{graphicx}

\makeatletter
\numberwithin{equation}{section}
\numberwithin{figure}{section}

\usepackage{algpseudocode,algorithm,algorithmicx}
\usepackage{tikz}
\usetikzlibrary{shapes,arrows,calc,decorations.pathreplacing}

\@ifundefined{showcaptionsetup}{}{%
 \PassOptionsToPackage{caption=false}{subfig}}
\usepackage{subfig}
\makeatother

\usepackage{babel}
\begin{document}

\title{A note on reducing spurious pressure oscillations in fully conservative
discontinuous Galerkin simulations of multicomponent flows}

\author{Eric J. Ching, Ryan F. Johnson, and Andrew D. Kercher}
\address{Laboratories for Computational Physics and Fluid Dynamics,  U.S. Naval Research Laboratory, 4555 Overlook Ave SW, Washington, DC 20375}

\begin{keyword}
Discontinuous Galerkin method; pressure equilibrium; spurious pressure
oscillations; contact discontinuity; contact interface
\end{keyword}
\maketitle
\global\long\def\middlebar{\,\middle|\,}%
\global\long\def\average#1{\left\{  \!\!\left\{  #1\right\}  \!\!\right\}  }%
\global\long\def\expnumber#1#2{{#1}\mathrm{e}{#2}}%
 \newcommand*{\horzbar}{\rule[.5ex]{2.5ex}{0.5pt}}

\global\long\def\revisionmath#1{\textcolor{red}{#1}}%

\makeatletter \def\ps@pprintTitle{  \let\@oddhead\@empty  \let\@evenhead\@empty  \def\@oddfoot{\centerline{\thepage}}  \let\@evenfoot\@oddfoot} \makeatother

\let\svthefootnote\thefootnote\let\thefootnote\relax\footnotetext{\\ \hspace*{65pt}DISTRIBUTION STATEMENT A. Approved for public release. Distribution is unlimited.}\addtocounter{footnote}{-1}\let\thefootnote\svthefootnote

\section{Introduction\label{sec:Introduction}}

The compressible, inviscid, multicomponent Euler equations in $d$
spatial dimensions are given by
\begin{equation}
\frac{\partial y}{\partial t}+\nabla\cdot\mathcal{F}\left(y\right)=0\label{eq:conservation-law-strong-form}
\end{equation}
where $t$ is the time, $y(x,t):\mathbb{R}^{d}\times\mathbb{R}^{+}\rightarrow\mathbb{R}^{m}$
is the vector of $m$ conservative variables, expanded as 

\begin{equation}
y=\left(\rho v_{1},\ldots,\rho v_{d},\rho e_{t},C_{1},\ldots,C_{n_{s}}\right)^{T},\label{eq:reacting-navier-stokes-state}
\end{equation}
and $\mathcal{F}(y):\mathbb{R}^{m}\rightarrow\mathbb{R}^{m\times d}$
is the convective flux, the $k$th spatial component of which is written
as 
\begin{equation}
\mathcal{F}_{k}\left(y\right)=\left(\rho v_{k}v_{1}+P\delta_{k1},\ldots,\rho v_{k}v_{d}+P\delta_{kd},v_{k}\left(\rho e_{t}+P\right),v_{k}C_{1},\ldots,v_{k}C_{n_{s}}\right)^{T}.\label{eq:reacting-navier-stokes-spatial-convective-flux-component}
\end{equation}
$x=\left(x_{1},\ldots,x_{d}\right)$ is the vector of physical coordinates,
$\rho$ is the density, $v=\left(v_{1},\ldots,v_{d}\right)$ is the
velocity vector, $e_{t}$ is the mass-specific total energy, $C=\left(C_{1},\ldots,C_{n_{s}}\right)$
is the vector of $n_{s}$ species concentrations, and $P$ is the
pressure, which is computed using the ideal gas law as
\begin{equation}
P=R^{0}T\sum_{i=1}^{n_{s}}C_{i},\label{eq:EOS}
\end{equation}
 where $R^{0}$ is the universal gas constant and $T$ is the temperature.
The mass fraction of the $i$th species is computed as
\[
Y_{i}=\frac{\rho_{i}}{\rho},
\]
where $\rho_{i}=W_{i}C_{i}$ is the partial density, with $W_{i}$
denoting the molecular weight of the $i$th species, and $\rho=\sum_{i=1}^{n_{s}}W_{i}C_{i}$
is the total density. In this work, we assume $d=1$ and mixtures
of thermally perfect gases, such that the specific heat capacities
vary with temperature.

It is well-known that spurious pressure oscillations are typically
generated at contact interfaces when using fully conservative numerical
schemes to solve Equation~(\ref{eq:conservation-law-strong-form})~\citep{Abg88,Kar94,Abg96},
an issue not observed when considering a monocomponent calorically
perfect gas. These pressure oscillations may lead to solver failure
and can still occur even if discrete entropy stability is satisfied~\citep{Gou20_2}.
Quasi-conservative methods are usually employed to prevent the occurrence
of such oscillations. For example, the double-flux approach~\citep{Abg01}
employs elementwise-constant, thermodynamically frozen auxiliary variables
to mimic a calorically perfect gas and mathematically guarantee preservation
of pressure equilibrium at contact interfaces. 

A popular numerical method that has successfully been applied to multicomponent
flows (both nonreacting and reacting) in recent years is the discontinuous
Galerkin (DG) method~\citep{Ree73,Bas97_2,Coc00}. This family of
numerical schemes boasts a number of desirable properties, such as
arbitrarily high order of accuracy on unstructured meshes, compatibility
with local grid and polynomial adaptation, and suitability for modern
compute hardware. Initial efforts to extend DG schemes to multicomponent
flows employed the double-flux approach~\citep{Bil11,Lv15,Ban20}.
This work, however, focuses on fully conservative DG schemes so that
any issues associated with not maintaining conservation are circumvented.
A handful of fully conservative DG schemes have been introduced that
can maintain pressure equilibrium at contact interfaces (while preserving
order of accuracy in smooth regions of the flow and without artificial
dissipation) in an approximate sense, i.e., pressure oscillations
can occur but remain small over long periods of time and do not cause
the solver to diverge if the solution is adequately resolved. For
example, Johnson and Kercher~\citep{Joh20_2} computed a canonical
test case involving the advection of a constant-pressure, hydrogen/oxygen
thermal bubble. They found that a colocated scheme successfully maintained
pressure equilibrium, but overintegration, which is important for
reducing aliasing errors, rapidly led to solver divergence. As such,
they developed an overintegration approach in which the flux is evaluated
with a modified state based on an approximate pressure that is projected
onto the selected finite element space via interpolation. This approach
was used to successfully compute complex multicomponent flows, including
a moving detonation wave and a reacting shear layer configuration~\citep{Joh20_2}.
Additionally, it can maintain pressure equilibrium when using multidimensional,
curved elements~\citep{Chi22_2} and can be easily incorporated into
a positivity-preserving, entropy-bounded framework~\citep{Chi22,Chi22_2}.
Bando~\citep{Ban23} further investigated this flux-evaluation approach
and compared it to another approach in which the pressure is projected
onto the finite element space via $L^{2}$-projection (as opposed
to interpolation). He found that in the same thermal-bubble test case,
the interpolation-based approach is simpler and yielded smaller deviations
from pressure equilibrium although both approaches (unlike standard
flux evaluation) maintained stability over long times. He also investigated
the influences of initialization strategy and choice of quadrature
rule. A somewhat similar approach was proposed by Franchina et al.~\citep{Fra16},
in which the primitive variables are treated as the unknowns and an
$L^{2}$-projection of the pressure is employed. Implicit time integration
was used in~\citep{Fra16}; though compatible with explicit time
stepping, a solution-dependent Jacobian term must be inverted at each
iteration, making it arguably less appropriate for explicit time integration
(which was used in~\citep{Joh20_2} and~\citep{Ban23}).

In this short note, we revisit the flux-evaluation approaches considered
in~\citep{Joh20_2} and~\citep{Ban23}, which successfully maintained
pressure equilibrium in the canonical hydrogen/oxygen thermal-bubble
advection test. We consider high-pressure, nitrogen/n-dodecane thermal-bubble
advection at low and high velocities. This type of test case has been
previously used to assess numerical schemes designed for simulating
transcritical/supercritical, real-fluid flows with, for example, cubic
equations of state and more complicated thermodynamic relations~\citep{Ma17,Boy21}.
Here, although we restrict ourselves to the thermally perfect gas
model, the increased nonlinearity of the thermodynamics of the nitrogen/n-dodecane
mixture is nevertheless perhaps more representative of certain complicated
mixtures at realistic conditions than the simpler hydrogen/oxygen
case. Furthermore, it makes the considered case more effective at
revealing deficiencies in numerical techniques. Also of interest is
the influence of projecting additional variables (other than solely
pressure) to the finite element trial space.

\section{Mathematical formulation}

\subsection{Discontinuous Galerkin discretization\label{subsec:DG-discretization}}

The semi-discrete form of Equation~(\ref{eq:conservation-law-strong-form})
is given as: find $y\in V_{h}^{p}$ such that
\begin{gather}
\sum_{\kappa\in\mathcal{T}}\left(\frac{\partial y}{\partial t},\mathfrak{v}\right)_{\kappa}-\sum_{\kappa\in\mathcal{T}}\left(\mathcal{F}\left(y\right),\nabla\mathfrak{v}\right)_{\kappa}+\sum_{\epsilon\in\mathcal{E}}\left(\mathcal{F}^{\dagger}\left(y^{+},y^{-},n\right),\left\llbracket \mathfrak{v}\right\rrbracket \right)_{\mathcal{E}}=0\qquad\forall\:\mathfrak{v}\in V_{h}^{p},\label{eq:semi-discrete-form}
\end{gather}
where $\text{\ensuremath{\left(\cdot,\cdot\right)}}$ denotes the
inner product, $\mathcal{T}$ is the set of cells $\kappa$, $\mathcal{E}$
is the set of interfaces $\epsilon$, $n$ is the normal vector at
a given interface, $y^{+}$ and $y^{-}$ are the states at both sides
of a given interface, $\mathcal{F}^{\dagger}\left(y^{+},y^{-},n\right)$
is the numerical flux, and $V_{h}^{p}$ is the space of basis and
test functions
\begin{eqnarray}
V_{h}^{p} & = & \left\{ \mathfrak{v}\in\left[L^{2}\left(\Omega\right)\right]^{m}\middlebar\forall\kappa\in\mathcal{T},\left.\mathfrak{v}\right|_{\kappa}\in\left[\mathsf{P}_{p}(\kappa)\right]^{m}\right\} ,\label{eq:discrete-subspace}
\end{eqnarray}
with $\mathsf{P}_{p}\left(\kappa\right)$ denoting the space of polynomial
functions of degree no greater than $p$ in $\kappa$. Only periodic
boundary conditions are considered in this work, such that all interfaces
are interior interfaces and the jump operator, $\left\llbracket \cdot\right\rrbracket $,
is defined as $\left\llbracket \cdot\right\rrbracket =\left(\cdot\right)^{+}-\left(\cdot\right)^{-}$.
We employ a nodal basis using Gauss-Lobatto points, such that the
element-local solution approximation is given by
\begin{equation}
y_{\kappa}=\sum_{j=1}^{n_{b}}y_{\kappa}(x_{j})\phi_{j},\label{eq:solution-approximation}
\end{equation}
where $n_{b}$ is the number of basis functions, $\left\{ \phi_{1},\ldots,\phi_{n_{b}}\right\} $
is the set of basis functions, and $\left\{ x_{1},\ldots,x_{n_{b}}\right\} $
is the set of node coordinates. The flux in Equation~(\ref{eq:semi-discrete-form})
is approximated as
\begin{equation}
\mathcal{F_{\kappa}}\approx\sum_{k=1}^{n_{c}}\mathcal{F}\left(\left.\text{\ensuremath{\mathcal{P}}}\left(z\left(y_{\kappa}\right)\right)\right|_{x_{k}}\right)\varphi_{k},\label{eq:flux-projection}
\end{equation}
where $n_{c}\geq n_{b}$ (with $n_{c}>n_{b}$ corresponding to overintegration),
$\left\{ \varphi_{1},\ldots,\varphi_{n_{c}}\right\} $ is the corresponding
set of basis functions (also based on Gauss-Lobatto points), $\mathcal{P}$
is a projection operator, and $z\left(y\right):\mathbb{R}^{m}\rightarrow\mathbb{R}^{m}$
is a vector of intermediate state variables. A major focus of this
study is assessing the effects of various choices of $\mathcal{P}$
and $z$, which will be detailed in the next subsection. The integrals
in Equation~(\ref{eq:semi-discrete-form}) are computed with a quadrature-free
approach~\citep{Atk96,Atk98} that is also employed in~\citep{Joh20_2}.
The overall trends observed here are expected to also apply to a more
conventional quadrature-based approach, which was employed in~\citep{Ban23};
indeed, in the context of maintaining pressure equilibrium in the
hydrogen/oxygen thermal-bubble configuration, both the quadrature-free
and quadrature-based approaches yielded very similar results.

\subsection{Flux evaluation}

This subsection discusses the choices of $\mathcal{P}$ and $z$ considered
in this study. 

\subsubsection{Projection operators}

As in~\citep{Ban23}, three projection operators are considered:
\begin{itemize}
\item $\mathcal{P}_{1}\left(\cdot\right)=\mathrm{id}\left(\cdot\right)$,
the identity function. This corresponds to a standard flux evaluation,
such that the RHS of Equation~(\ref{eq:flux-projection}) reduces
to $\sum_{k=1}^{n_{c}}\mathcal{F}\left(y_{\kappa}\left(x_{k}\right)\right)\varphi_{k}$,
regardless of the choice of $z$.
\item $\mathcal{P}_{2}\left(\cdot\right)=\mathcal{I}\left(\cdot\right)$,
interpolatory projection onto $V_{h}^{p}$. Note that using a colocated
scheme, this is equivalent to the identity function. This approach
was first introduced by Johnson and Kercher~\citep{Joh20_2}.
\item $\mathcal{P}_{3}\left(\cdot\right)=\Pi\left(\cdot\right)$, $L^{2}$-projection
onto $V_{h}^{p}$.
\end{itemize}
In this study, non-identity projection is applied only when overintegration
is employed. Optimal convergence was previously observed when using
interpolatory projection, $\mathcal{P}_{2}$,~\citep{Joh20_2,Ban23}
and $L^{2}$-projection, $\mathcal{P}_{3}$,~\citep{Ban23}; therefore,
we do not assess order of convergence here.

\subsubsection{Intermediate state variables}

Three choices of $z$ are considered:
\begin{itemize}
\item $z_{1}=\left(\rho v_{1},\ldots,\rho v_{d},P,C_{1},\ldots,C_{n_{s}}\right)^{T}$,
where total energy is replaced with pressure, which was proposed by
Johnson and Kercher~\citep{Joh20_2}. Note that this was the only
choice of intermediate state variables considered in~\citep{Joh20_2}
and~\citep{Ban23}. 
\item $z_{2}=\left(v_{1},\ldots,v_{d},P,C_{1},\ldots,C_{n_{s}}\right)^{T}$,
where total energy is replaced with pressure and momentum is replaced
with velocity.
\item $z_{3}=\left(v_{1},\ldots,v_{d},P,T,Y_{1},\ldots,Y_{n_{s}-1}\right)^{T}$,
which represents a full set of primitive variables.
\end{itemize}

\subsubsection{Integration}

We compute $p=2$ solutions using both colocated integration and overintegration.
The former was found to maintain stability over long times in the
hydrogen-oxygen thermal-bubble advection test~\citep{Joh20_2,Ban23},
whereas the latter, in conjunction with $\mathcal{P}_{1}$ (i.e.,
standard flux evaluation), rapidly led to solver divergence. For colocated
integration, only $\mathcal{P}_{1}$ is applied, which corresponds
to standard colocated integration. For overintegration, the 5-point
Gauss-Lobatto nodal set is used for the integration points. Note that
Bando~\citep{Ban23} assessed the effects of different quadrature
rules. Such an investigation is not repeated here.

\section{Results}

This section presents results for the advection of a high-pressure,
nitrogen/n-dodecane thermal bubble~\citep{Ma17,Boy21}. Although
this test case was presented in~\citep{Ma17} and~\citep{Boy21}
with discontinuous initial conditions, discontinuity-capturing schemes
were applied in those studies. Here, we are specifically interested
in the performance of DG schemes without any discontinuity-capturing
techniques (e.g., dissipative limiters or artificial viscosity). Similar
to the hydrogen/oxygen thermal-bubble configuration considered in~\citep{Lv15,Joh20_2,Ban23},
we modify the initial condition to be smooth (but still with high
gradients) as
\begin{eqnarray}
Y_{n\text{-}\mathrm{C_{12}H_{26}}} & = & \frac{1}{2}\left[1-\tanh\left(25|x|-0.2\right)\right],\nonumber \\
Y_{\mathrm{N_{2}}} & = & 1-Y_{n\text{-}\mathrm{C_{12}H_{26}}},\label{eq:thermal-bubble}\\
T & = & \frac{T_{\min}+T_{\max}}{2}-\frac{T_{\max}-T_{\min}}{2}\tanh\left(25|x|-0.2\right)\textrm{ K},\nonumber \\
P & = & 6\textrm{ MPa},\nonumber 
\end{eqnarray}
where $T_{\min}=363\text{ K}$ and $T_{\max}=900\text{ K}$. In the
hydrogen/oxygen case, despite the smooth initial condition, underresolution
leads to spurious pressure oscillations that, with standard overintegration,
grow rapidly and lead to solver divergence~\citep{Joh20_2,Ban23}.
We consider two advection velocities, $v=1\text{ m/s}$, as in~\citep{Ma17},
and $v=600\text{ m/s}$, as in~\citep{Boy21}, since we have found
that changing the velocity can yield markedly different results. The
computational domain is $\left[-0.5,0.5\right]\text{ m}$, partitioned
into $50$ line cells. Both sides of the domain are periodic. The
HLLC~\citep{Tor13} flux function is employed. All solutions are
initialized using interpolation and then integrated forward in time
for ten advection periods using the third-order strong-stability-preserving
Runge-Kutta scheme~\citep{Got01} with $\mathrm{CFL}=0.8$ based
on the order-dependent linear-stability constraint. To assess deviations
from pressure equilibrium, we compute the following global quantity
as a function of time~\citep{Ban23}:
\begin{equation}
\Delta P(t)=\max_{x}P(t,x)-\min_{x}P(t,x),\label{eq:DeltaP}
\end{equation}
where the maximum and minimum are evaluated over all solution nodes.
All simulations are performed using a modified version of the JENRE\textregistered~Multiphysics
Framework~\citep{Cor18_SCITECH,Joh20_2} that incorporates the techniques
described in this work. In the following, $\tau$ denotes the time
to advect the solution one period, and $P_{0}$ is the initial pressure
of $6\text{ MPa}$.

\subsection{Low velocity, $v=1\text{ m/s}$\label{subsec:low-velocity}}

\begin{figure}[H]
\subfloat[\label{fig:thermal_bubble_dodecane_deltaP_1_m-s}Temporal variation
of $\Delta P$.]{\includegraphics[width=0.48\columnwidth]{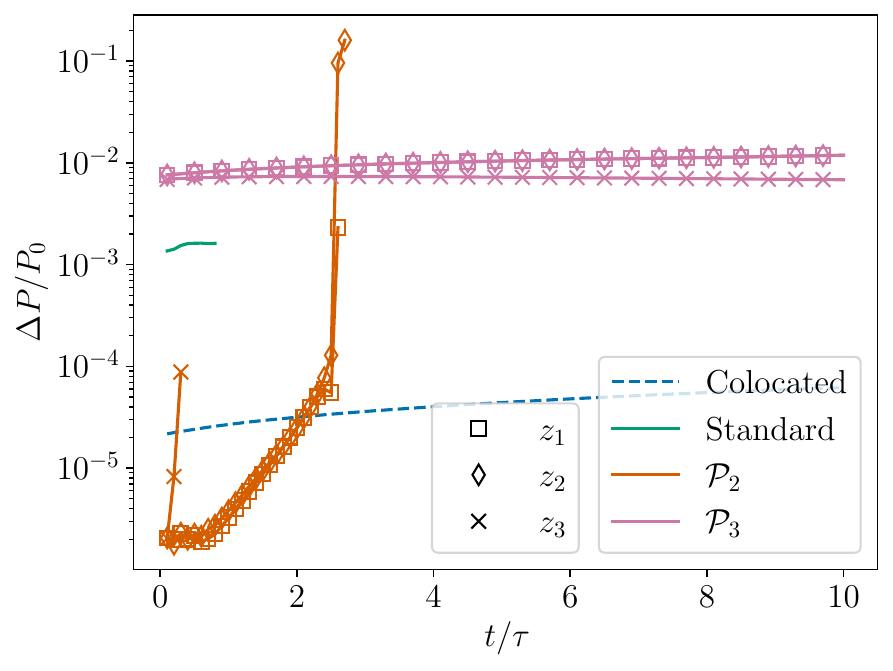}}\hfill{}\subfloat[\label{fig:thermal_bubble_dodecane_T_1_m-s}Temperature profiles at
$t=10\tau$.]{\includegraphics[width=0.48\columnwidth]{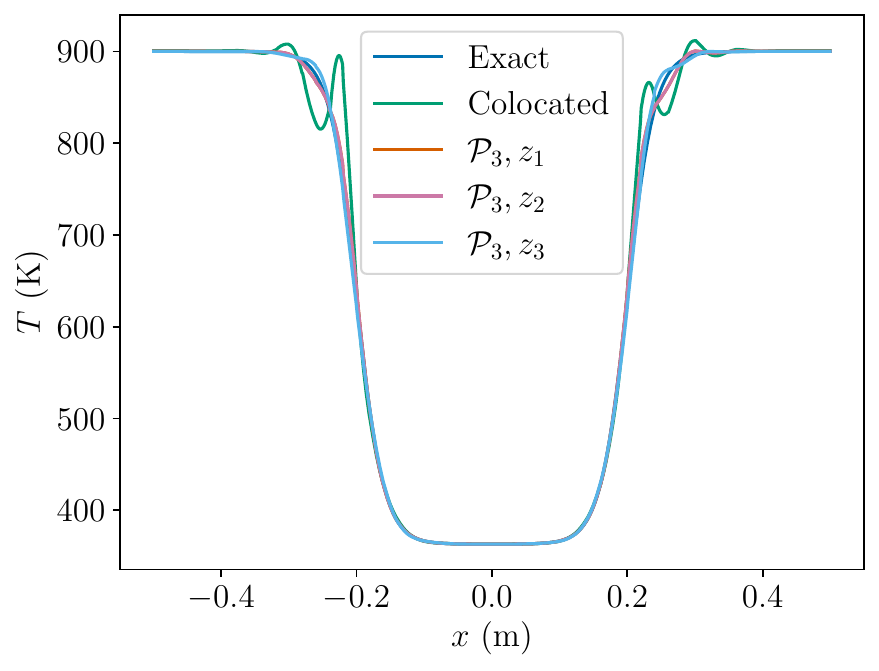}}

\caption{\label{fig:thermal_bubble_dodecane_1_m-s}Solution to the advection
of a nitrogen/n-dodecane thermal bubble at $v=1\text{ m/s}$. Figure~\ref{fig:thermal_bubble_dodecane_deltaP_1_m-s}
shows the temporal variation of $\Delta P$, sampled at intervals
of $\tau/10$ (for a total of 100 samples if the solution does not
diverge). For the $\mathcal{P}_{3}$ results, only every fourth marker
is shown. Figure~\ref{fig:thermal_bubble_dodecane_T_1_m-s} displays
the temperature profiles at $t=10\tau$ for the stable solutions.}
\end{figure}

Figure~\ref{fig:thermal_bubble_dodecane_deltaP_1_m-s} presents the
temporal variation of $\Delta P$ for the following approaches:
\begin{itemize}
\item Colocated integration.
\item Standard overintegration, i.e., $\mathcal{P}_{1}$.
\item Modified overintegration using all combinations of $\left\{ \mathcal{P}_{2},\mathcal{P}_{3}\right\} $
and $\left\{ z_{1},z_{2},z_{3}\right\} $.
\end{itemize}
$\Delta P$ is sampled every $\tau/10$ seconds until the solution
either diverges or is advected for ten periods (i.e., $t=10\tau$).
The solution diverges rapidly in the case of standard overintegration
or modified integration with $\left\{ \mathcal{P}_{2},z_{3}\right\} $.
$\mathcal{P}_{2}$ in conjunction with $z_{2}$ or $z_{3}$ maintains
solution stability for longer times, although the solution nevertheless
blows up well before $t=10\tau$. Only colocated integration and $\mathcal{P}_{3}$,
regardless of choice of $z$, maintain stability through ten advection
periods. Deviations from pressure equilibrium are overall smallest
with colocated integration. In the context of $\mathcal{P}_{3}$,
$z_{3}$ preserves pressure equilibrium slightly better than $z_{1}$
and $z_{2}$, both of which give nearly identical results. At early
times, however, $\mathcal{P}_{2}$ yields the smallest pressure oscillations
(which then grow rapidly before leading to solver failure). Figure~\ref{fig:thermal_bubble_dodecane_T_1_m-s}
presents the temperature profiles for the stable solutions at $t=10\tau$.
Despite most effectively maintaining pressure equilibrium, colocated
integration leads to appreciable oscillations in the temperature profile
that are not observed in the $\mathcal{P}_{3}$ solutions, all of
which exhibit good agreement with the exact solution. 

\subsection{High velocity, $v=600\text{ m/s}$}

\begin{figure}[H]
\subfloat[\label{fig:thermal_bubble_dodecane_deltaP_600_m-s}Temporal variation
of $\Delta P$.]{\includegraphics[width=0.48\columnwidth]{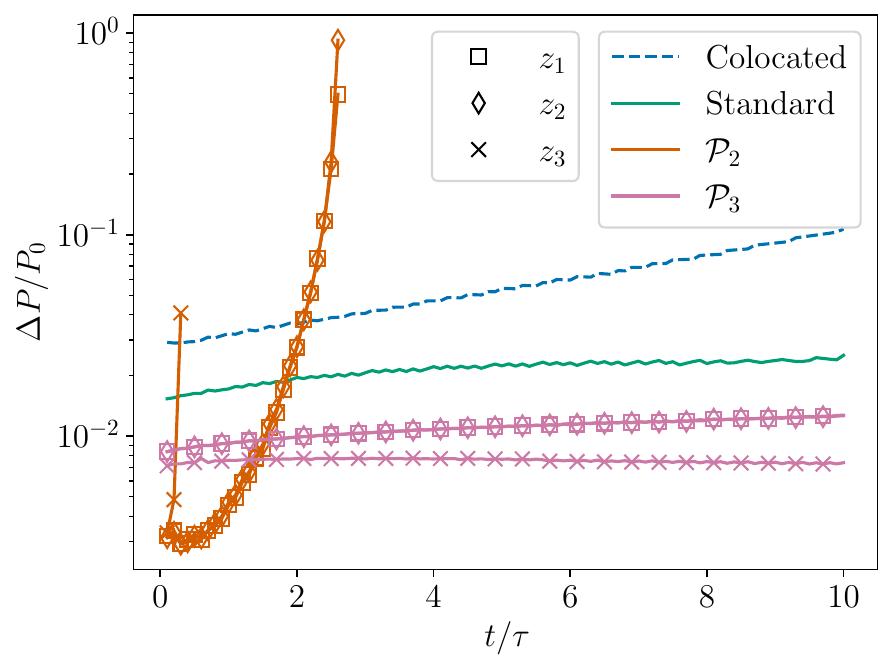}}\hfill{}\subfloat[\label{fig:thermal_bubble_dodecane_T_600_m-s}Temperature profiles
at $t=10\tau$.]{\includegraphics[width=0.48\columnwidth]{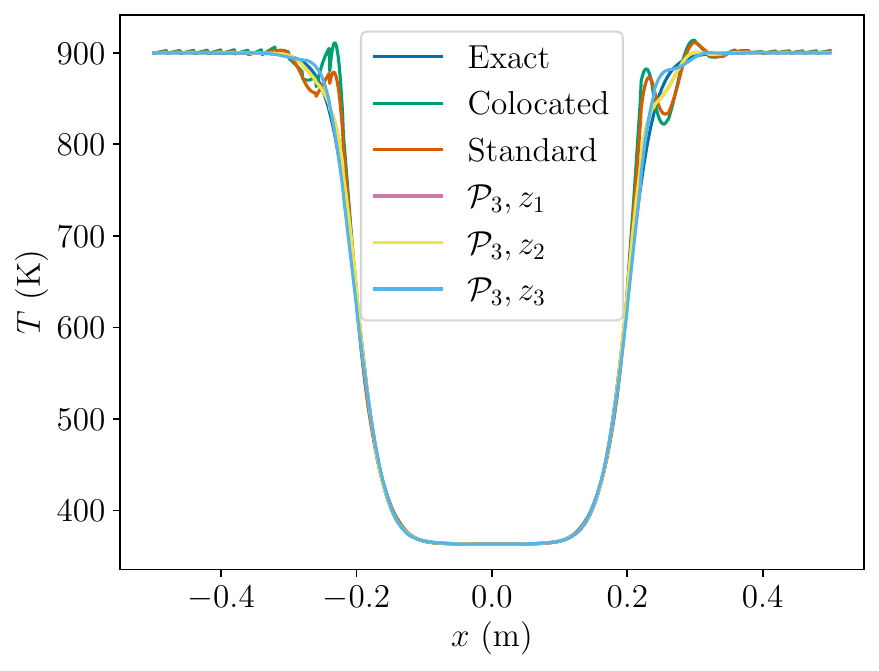}}

\caption{\label{fig:thermal_bubble_dodecane_600_m-s}Solution to the advection
of a nitrogen/n-dodecane thermal bubble at $v=600\text{ m/s}$. Figure~\ref{fig:thermal_bubble_dodecane_deltaP_600_m-s}
shows the temporal variation of $\Delta P$, sampled at intervals
of $\tau/10$ (for a total of 100 samples if the solution does not
diverge). For the $\mathcal{P}_{3}$ results, only every fourth marker
is shown. Figure~\ref{fig:thermal_bubble_dodecane_T_600_m-s} displays
the temperature profiles at $t=10\tau$ for the stable solutions.}
\end{figure}

As in the previous subsection, Figure~\ref{fig:thermal_bubble_dodecane_deltaP_600_m-s}
displays the temporal variation of $\Delta P$ for the considered
approaches. The $\mathcal{P}_{2}$ solutions again initially exhibit
very small pressure oscillations before diverging, with the $z_{3}$
case diverging most rapidly. Furthermore, when using $\mathcal{P}_{3}$,
$z_{3}$ again gives smaller deviations from pressure equilibrium
than $z_{1}$ and $z_{2}$. However, key differences with the low-velocity
case are observed. First, the magnitude of pressure oscillations is
generally greater in the high-velocity case, although the $\mathcal{P}_{3}$
solutions yield deviations of similar magnitude between both velocities.
Second, standard overintegration maintains stability throughout the
simulation. Additionally, of the stable solutions, the colocated case
exhibits the largest pressure deviations, which would likely continue
to grow if the simulation is run for longer times. In contrast, the
maximum pressure deviations for the other stable solutions seem to
have either plateaued or begun to plateau. Pressure equilibrium is
the most well-preserved in the $\mathcal{P}_{3}$ solutions, particularly
in conjunction with $z_{3}$. 

Figure~\ref{fig:thermal_bubble_dodecane_T_600_m-s} presents the
temperature profiles of the stable solutions at $t=10\tau$. Temperature
oscillations are observed in the cases of colocated integration and
standard overintegration. In contrast, the $\mathcal{P}_{3}$ solutions
exhibit better agreement with the exact temperature.

\begin{figure}[H]
\begin{centering}
\includegraphics[width=0.48\columnwidth]{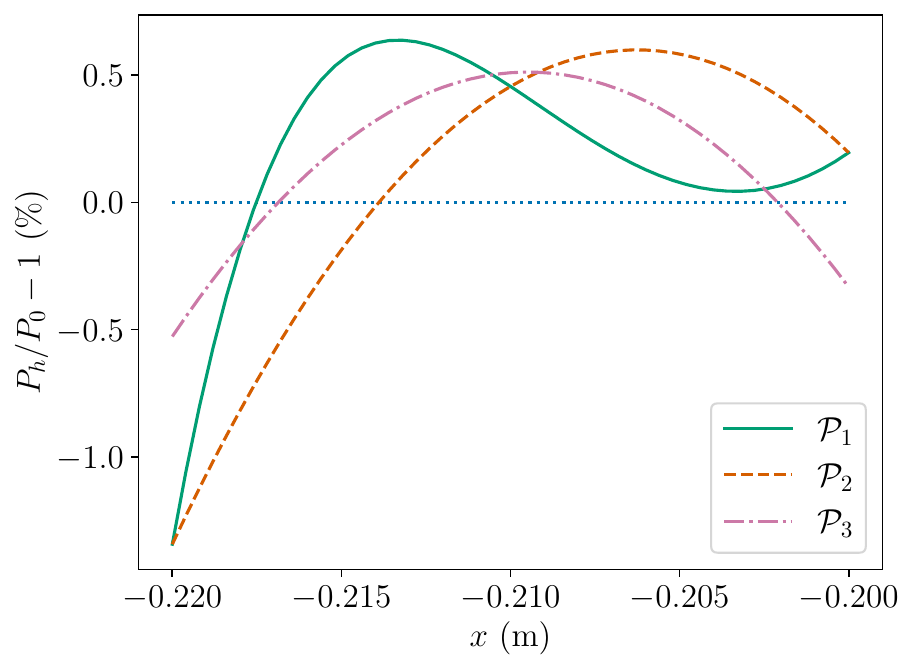}
\par\end{centering}
\caption{\label{fig:thermal_bubble_dodecane_pressure_error_600m-s}Pressure-deviation
profiles in the cell $\kappa=\left[-0.22,-0.2\right]\text{ m}$ evaluated
with the $\mathcal{P}_{1}$, $\mathcal{P}_{2}$, and $\mathcal{P}_{3}$
projection operators for the $\mathcal{P}_{1}$ solution at $t=10\tau$.
The dotted blue line corresponds to exact pressure equilibrium. }
\end{figure}

To help explain why overintegration with $\mathcal{P}_{3}$ can lead
to higher solution stability than both standard overintegration (i.e.,
$\mathcal{P}_{1}$) and overintegration with $\mathcal{P}_{2}$, we
take the $\mathcal{P}_{1}$ solution at $t=10\tau$ and then evaluate
the pressure in the cell $\kappa=\left[-0.22,-0.2\right]\text{ m}$
with the $\mathcal{P}_{1}$, $\mathcal{P}_{2}$, and $\mathcal{P}_{3}$
projection operators. The resulting pressure-deviation profiles in
the given cell are displayed in Figure~\ref{fig:thermal_bubble_dodecane_pressure_error_600m-s}.
Noticeable pressure deviations are located at the faces in the $\mathcal{P}_{1}$
case (i.e., the pressure is evaluated normally); since $\mathcal{P}_{2}$
(at least using Gauss-Lobatto points) does not modify pressure at
the faces, overall deviation from pressure equilibrium is not appreciably
reduced. In contrast, $\mathcal{P}_{3}$ markedly improves preservation
of pressure equilibrium throughout the cell, indicating that $L^{2}$-projection
can act as a mechanism to reduce spurious oscillations that initially
emerge via, for instance, underresolution or inexact evaluation of
the flux. In the case of $z_{2}$ or $z_{3}$, oscillations of other
variables can also be mitigated. However, it should be noted that
$\mathcal{P}_{2}$ can reduce pressure oscillations if deviations
are large \emph{inside} the cell but not at the faces. Furthermore,
there is no guarantee that $\mathcal{P}_{3}$ will always reduce spurious
oscillations.

\section{Concluding remarks}

In this short note, we reevaluated previously introduced techniques
designed to reduce spurious pressure oscillations at contact interfaces
in multicomponent flows in the context of overintegrated DG discretizations~\citep{Joh20_2,Ban23}.
Specifically, we focused on strategies that do not (a) introduce conservation
error, (b) rely on artificial viscosity or limiting, or (c) degrade
order of accuracy in smooth regions of the flow. The considered techniques,
which employ a projection of the pressure (and potentially additional
variables) onto the finite element trial space via either interpolation
or $L^{2}$-projection, were previously shown to maintain approximate
pressure equilibrium and thus stable solutions over long times (in
contrast to a standard DG scheme with overintegration) in the advection
of a hydrogen/oxygen thermal bubble. In~\citep{Joh20_2}, interpolation-based
projection was considered the preferred approach due to better preservation
of pressure equilibrium.

In this work, we considered a more challenging test case: advection
of a high-pressure nitrogen/n-dodecane thermal bubble at both low
and high velocities. All simulations were performed on a 50-cell grid
using $p=2$. Key observations from this test case are as follows:
\begin{itemize}
\item Interpolation-based projection always led to solver divergence. Although
it overall maintained stability for longer times than standard overintegration
in the low-velocity case, it was outperformed by standard overintegration
in the high-velocity case. 
\item $L^{2}$-projection always maintained solution stability. It produced
larger pressure deviations than colocated integration in the low-velocity
case but more effectively maintained pressure equilibrium than all
other approaches in the high-velocity case. 
\item Performing $L^{2}$-projection of a full set of primitive variables
(specifically, mass fractions, pressure, temperature, and velocity)
more effectively preserved pressure equilibrium than if only pressure
or if both pressure and velocity were projected. However, projecting
a smaller set of variables may be sufficient for stability.
\item $L^{2}$-projection led to better predictions of temperature than
colocated integration and standard overintegration.
\item Colocated integration always maintained solution stability, but led
to inferior temperature predictions, indicating that higher resolution
(than in the case of overintegration) may be needed to offset the
greater integration error.
\end{itemize}
These findings suggest that this configuration is more effective at
testing the ability of a numerical scheme to preserve pressure equilibrium,
even without consideration of a cubic equation of state or more complicated
thermodynamic relations commonly associated with it~\citep{Ma17,Boy21}.
Furthermore, it is also valuable to consider different advection velocities.
It should be also be noted that although the $L^{2}$-projection-based
approach is the only overintegration strategy that prevented solver
divergence across all considered conditions, the interpolation-based
approach is simpler and was shown to maintain robustness across a
variety of challenging test cases involving more realistic flow conditions
and geometries, including moving detonation waves and a chemically
reacting shear layer~\citep{Joh20_2}, suggesting that it may still
be a reliable choice. 

Future work will involve consideration of real-fluid effects, in particular
a cubic equation of state and thermodynamic departure functions~\citep{Ma17},
in multiple dimensions. The ability of the considered techniques,
specifically the $L^{2}$-projection-based strategy, will be further
assessed in this more challenging context of transcritical and supercritical
flows. Furthermore, a fully conservative finite volume scheme that
mathematically guarantees preservation of pressure equilibrium was
recently developed by Fujiwara et al.~\citep{Fuj23}; an extension
to DG schemes may indeed be worth pursuing.

\section*{Acknowledgments}

This work is sponsored by the Office of Naval Research through the
Naval Research Laboratory 6.1 Computational Physics Task Area. 

\bibliographystyle{elsarticle-num}
\bibliography{../JCP_submission/citations}

\begin{thebibliography}{10}
\expandafter\ifx\csname url\endcsname\relax
  \def\url#1{\texttt{#1}}\fi
\expandafter\ifx\csname urlprefix\endcsname\relax\def\urlprefix{URL }\fi
\expandafter\ifx\csname href\endcsname\relax
  \def\href#1#2{#2} \def\path#1{#1}\fi

\bibitem{Abg88}
R.~Abgrall, Generalisation of the {R}oe scheme for the computation of mixture
  of perfect gases, La Recherche Aérospatiale 6 (1988) 31--43.

\bibitem{Kar94}
S.~Karni, Multicomponent flow calculations by a consistent primitive algorithm,
  Journal of Computational Physics 112~(1) (1994) 31 -- 43.
\newblock \href {https://doi.org/https://doi.org/10.1006/jcph.1994.1080}
  {\path{doi:https://doi.org/10.1006/jcph.1994.1080}}.

\bibitem{Abg96}
R.~Abgrall, How to prevent pressure oscillations in multicomponent flow
  calculations: A quasi conservative approach, Journal of Computational Physics
  125~(1) (1996) 150 -- 160.
\newblock \href {https://doi.org/https://doi.org/10.1006/jcph.1996.0085}
  {\path{doi:https://doi.org/10.1006/jcph.1996.0085}}.

\bibitem{Gou20_2}
A.~Gouasmi, K.~Duraisamy, S.~M. Murman, Formulation of entropy-stable schemes
  for the multicomponent compressible {E}uler equations, Computer Methods in
  Applied Mechanics and Engineering 363 (2020) 112912.

\bibitem{Abg01}
R.~Abgrall, S.~Karni, Computations of compressible multifluids, Journal of
  Computational Physics 169~(2) (2001) 594 -- 623.
\newblock \href {https://doi.org/https://doi.org/10.1006/jcph.2000.6685}
  {\path{doi:https://doi.org/10.1006/jcph.2000.6685}}.

\bibitem{Ree73}
W.~H. Reed, T.~Hill, Triangular mesh methods for the neutron transport
  equation, Tech. rep., Los Alamos Scientific Lab., N. Mex.(USA) (1973).

\bibitem{Bas97_2}
F.~Bassi, S.~Rebay, High-order accurate discontinuous finite element solution
  of the 2{D} {E}uler equations, Journal of Computational Physics 138~(2)
  (1997) 251--285.

\bibitem{Coc00}
B.~Cockburn, G.~Karniadakis, C.-W. Shu, The development of discontinuous
  {G}alerkin methods, in: Discontinuous {G}alerkin Methods, Springer, 2000, pp.
  3--50.

\bibitem{Bil11}
G.~Billet, J.~Ryan, A {R}unge–{K}utta discontinuous {G}alerkin approach to
  solve reactive flows: The hyperbolic operator, Journal of Computational
  Physics 230~(4) (2011) 1064 -- 1083.
\newblock \href {https://doi.org/https://doi.org/10.1016/j.jcp.2010.10.025}
  {\path{doi:https://doi.org/10.1016/j.jcp.2010.10.025}}.

\bibitem{Lv15}
Y.~Lv, M.~Ihme, Discontinuous {G}alerkin method for multicomponent chemically
  reacting flows and combustion, Journal of Computational Physics 270 (2014)
  105 -- 137.
\newblock \href {https://doi.org/https://doi.org/10.1016/j.jcp.2014.03.029}
  {\path{doi:https://doi.org/10.1016/j.jcp.2014.03.029}}.

\bibitem{Ban20}
K.~Bando, M.~Sekachev, M.~Ihme, Comparison of algorithms for simulating
  multi-component reacting flows using high-order discontinuous {G}alerkin
  methods (2020).
\newblock \href {https://doi.org/10.2514/6.2020-1751}
  {\path{doi:10.2514/6.2020-1751}}.

\bibitem{Joh20_2}
R.~F. Johnson, A.~D. Kercher, A conservative discontinuous {G}alerkin
  discretization for the chemically reacting {N}avier--{S}tokes equations,
  Journal of Computational Physics 423 (2020) 109826.
\newblock \href {https://doi.org/10.1016/j.jcp.2020.109826}
  {\path{doi:10.1016/j.jcp.2020.109826}}.

\bibitem{Chi22_2}
E.~J. Ching, R.~F. Johnson, A.~D. Kercher, Positivity-preserving and
  entropy-bounded discontinuous {G}alerkin method for the chemically reacting,
  compressible {E}uler equations. {P}art {II}: {T}he multidimensional case,
  arXiv preprint arXiv:2211.16297~\url{https://arxiv.org/abs/2211.16297}
  (2022).

\bibitem{Chi22}
E.~J. Ching, R.~F. Johnson, A.~D. Kercher, Positivity-preserving and
  entropy-bounded discontinuous {G}alerkin method for the chemically reacting,
  compressible {E}uler equations. {P}art {I}: {T}he one-dimensional case, arXiv
  preprint arXiv:2211.16254~\url{https://arxiv.org/abs/2211.16254} (2022).

\bibitem{Ban23}
K.~Bando, Towards high-performance discontinuous {G}alerkin simulations of
  reacting flows using {L}egion, Ph.D. thesis, Stanford University (2023).

\bibitem{Fra16}
N.~Franchina, M.~Savini, F.~Bassi, Multicomponent gas flow computations by a
  discontinuous {G}alerkin scheme using {L}2-projection of perfect gas {EOS},
  Journal of Computational Physics 315 (2016) 302--322.

\bibitem{Ma17}
P.~C. Ma, Y.~Lv, M.~Ihme, An entropy-stable hybrid scheme for simulations of
  transcritical real-fluid flows, Journal of Computational Physics 340 (2017)
  330--357.

\bibitem{Boy21}
B.~Boyd, D.~Jarrahbashi, A diffuse-interface method for reducing spurious
  pressure oscillations in multicomponent transcritical flow simulations,
  Computers \& Fluids 222 (2021) 104924.

\bibitem{Atk96}
H.~Atkins, C.~Shu, Quadrature-free implementation of discontinuous {G}alerkin
  methods for hyperbolic equations, {ICASE} {R}eport 96-51, 1996, Tech. rep.,
  NASA Langley Research Center, nASA-CR-201594 (August 1996).

\bibitem{Atk98}
H.~L. Atkins, C.-W. Shu, Quadrature-free implementation of discontinuous
  {G}alerkin method for hyperbolic equations, AIAA Journal 36~(5) (1998)
  775--782.

\bibitem{Tor13}
E.~Toro, Riemann solvers and numerical methods for fluid dynamics: {A}
  practical introduction, Springer Science \& Business Media, 2013.

\bibitem{Got01}
S.~Gottlieb, C.~Shu, E.~Tadmor, Strong stability-preserving high-order time
  discretization methods, SIAM review 43~(1) (2001) 89--112.

\bibitem{Cor18_SCITECH}
A.~Corrigan, A.~Kercher, J.~Liu, K.~Kailasanath, Jet noise simulation using a
  higher-order discontinuous {G}alerkin method, in: 2018 AIAA SciTech Forum,
  2018, {A}IAA-2018-1247.

\bibitem{Fuj23}
Y.~Fujiwara, Y.~Tamaki, S.~Kawai, Fully conservative and pressure-equilibrium
  preserving scheme for compressible multi-component flows, Journal of
  Computational Physics 478 (2023) 111973.

\end{thebibliography}

\end{document}